\documentclass[aps,pra,10pt,twocolumn,superscriptaddress,floatfix,nofootinbib,showpacs,longbibliography]{revtex4-2}

\usepackage{physics}
\usepackage{amssymb}
\usepackage{graphicx}
\usepackage{gensymb}
\usepackage{float}
\usepackage{lipsum}
\usepackage{adjustbox}
\usepackage{graphicx} % Required for inserting images
\usepackage{xcolor}
\usepackage[colorlinks=true,linkcolor=blue,urlcolor=magenta,citecolor=red]{hyperref} 
\usepackage{caption}
\usepackage{subcaption}
\usepackage{amsfonts} 

\graphicspath{ {./figures/} }

\begin{document}

%================================================================================
\title{Four-qubit photonic system for publicly verifiable quantum random\\ numbers and generation of public and private key}

%\title{Generation of publicly verifiable quantum random numbers, public and \\
%private key from a four-qubit photonic system}

\author{Mayalakshmi Kolangatt}
	\affiliation{Quantum Optics \& Quantum Information, Department of Instrumentation and Applied Physics, Indian Institute of Science, Bengaluru 560012, India.}
\author{Anirudh Verma}
	\affiliation{Quantum Optics \& Quantum Information, Department of Instrumentation and Applied Physics, Indian Institute of Science, Bengaluru 560012, India.}
\author{Sujai Matta}
	\affiliation{Quantum Optics \& Quantum Information, Department of Instrumentation and Applied Physics, Indian Institute of Science, Bengaluru 560012, India.}
\author{Kanad Sengupta}
	\affiliation{Quantum Optics \& Quantum Information, Department of Instrumentation and Applied Physics, Indian Institute of Science, Bengaluru 560012, India.}
%\author{Mayank Joshi}
%	\affiliation{Quantum Optics \& Quantum Information, Department of Instrumentation and Applied Physics, Indian Institute of Science, Bengaluru 560012, India.}
%\author{Muhammed K Shafi}
%	\affiliation{Quantum Optics \& Quantum Information, Department of Instrumentation and Applied Physics, Indian Institute of Science, Bengaluru 560012, India.}
		\author{C. M. Chandrashekar}
	\email{chandracm@iisc.ac.in}
	\affiliation{Quantum Optics \& Quantum Information, Department of Instrumentation and Applied Physics, Indian Institute of Science, Bengaluru 560012, India.}
	\affiliation{The Institute of Mathematical Sciences, C. I. T. Campus, Taramani, Chennai 600113, India}
	\affiliation{Homi Bhabha National Institute, Training School Complex, Anushakti Nagar, Mumbai 400094, India}

%================================================================================

%================================================================================

\begin{abstract}

\noindent

We theoretically propose and experimentally demonstrate the use of a configurable four-qubit photonic system to generate a publicly verifiable quantum random numbers, to perform entanglement verification, and to generate secure public and private key. Quantum circuits, to generate the desired four-qubit states and its experimental realization in the photonic architecture is carried out using photon pairs entangled in polarization and path degree of freedom.  By performing measurements on the four-qubit system and accessing partial information of the four-qubit state for public verification, we generate publicly verified and purely secured random bits at the rate of 185 kbps from collective data of 370 kbps. When the system is used for generating public and private keys, an equal number of public and private keys are generated simultaneously. We also record about 97.9\% of sampled bits from four-qubit states passing entanglement verification and demonstrate the use of public and private key generated for image encryption-decryption. The theoretical model of noise on the four-qubit state and its effect on the generation rate of verified and secured bits are in perfect agreement with the experimental results. This demonstrates the practical use of the small-scale multi-qubit photonic system for quantum-safe applications by providing the option for real-time verification of the security feature of the quantum system.
\end{abstract}
%================================================================================

\maketitle

%================================================================================
\section{Introduction}

%================================================================================
Randomness generators are a very crucial resource and we employ them readily in our everyday computational and communication activities, including cryptography\,\cite{SW2009, G2006}, scientific simulations, gaming, statistical sampling, and many others\,\cite{BM2007, DCL2019}. Most of the random number generators that are employed today for practical use are sourced from algorithms on a classical computer.  With a knowledge of the algorithm and the initial input, the random sequence can be vulnerable to prediction and hence they are not true random number generators (TRNGs). They are better classified as pseudo-random number generators (PRNG)\,\cite{J1990, WJG2004, HD1962}. Ideally, the output of the random number generator should adhere to the principles of uniform distribution and independence. Several statistical tests exist that can verify the randomness of the observed sequence\,\cite{M1992, K1998, RSN2001}. Despite passing all statistical tests, it is nearly impossible to distinguish between a random sequence from a PRNG and a genuinely random sequence, TRNGs. This is because true randomness cannot be generated and certified using classical methods. The inherent probabilistic nature of quantum physics enables the generation of true randomness from a quantum system\,\cite{S1970, BAK2017}. Hence, a quantum system is becoming a new standard of being an optimal source for generating genuine random numbers\,\cite{XXZ2016, MJ2017, MPJ2021}. Moreover, the principles of quantum mechanics can be applied to assess the quality of these random numbers. Quantum technologies have progressed to the point where this can be achieved. Different schemes of quantum random number generators (QRNGs) have been demonstrated using various quantum sources as resources for randomness\,\cite{MJ2017}. Entangled and single photons from the spontaneous parametric down-conversion (SPDC) process\,\cite{QMD2004, BB2010} and single-photon's entanglement states\,\cite{KCK2009, MLA2021, LAM2022, SCH2023} have also been used to generate quantum random numbers. Depending on the method and characterization of the devices used for QRNG, the protocols are broadly classified into the following categories, namely, trusted device, semi-device independent sources\,\cite{MJ2017} and device-independent sources\,\cite{AMV2018,CSA2023}. Current QRNGs are trusted devices, next-generation self-testing, and semi-self-testing QRNGs are under development, and selecting a QRNG method requires considering multiple factors based on user needs\,\cite{MMP2023}.

In addition to generating genuine random numbers, subjecting the random bit sequence to public verification is very important to ensure the trustworthiness of the device to the beneficiary. This would involve a third party accessing the bit string and performing a randomness test resulting in a compromise of the secrecy of the generated bit. Even though entangled photon pairs are used for generating certified quantum random numbers,  using one of the entangled photons for public verification of randomness will make it vulnerable to the secrecy of the bit generated by the other photon due to the quantum correlation between the photons. Recently, the scheme for using a multi-qubit entangled state which will allow public verification without compromising on the secrecy of the undisclosed bit has been proposed\,\cite{JWW2020, JJJ2024} and its experimental demonstration using entangled photon pairs in three-qubit configuration has been reported\,\cite{TAC2024}.

In this work, we present a configurable four-qubit photonic system in the framework of a quantum circuit. 
 A particular configuration of two photons entangled in polarization and path degree of freedom is generated using polarization-entangled state\,\cite{LKC2016} as the initial state is further configured to generate entanglement in path degree. We experimentally demonstrate the use of the four-qubit photonic system to generate publicly verifiable quantum random numbers without compromising on the secrecy of the undisclosed bits. The random numbers obtained from the scheme pass all the randomness tests in the NIST test suite\,\cite{RSN2001}.  We also show that different configurations of the four-qubit state can be prepared to demonstrate entanglement verification with access to sample space of all four strings generated and to two of the four strings of bits generated in real-time, respectively. In addition, we also show that the four-qubit entangled state generated can serve as two entangled subsystems with access of one photon to each subsystem and use it to simultaneously generate private and public keys for secure communication between the two subsystem.  Theoretical model of noise on the four-qubit state and its effect on the generation rate of verified and secured bits are numerically analyzed and shown to be in perfect agreement with the experimental results. This demonstrates the practical use of small-scale photonic quantum computers for multiple quantum-safe applications by providing the option for real-time verification of the genuine and verifiable security feature of the multi-qubit quantum system.

%=========================================
\section{Theoretical Description}\label{Theory}
%=========================================

\subsection{Publicly verifiable quantum random numbers}

A state from the Hilbert space $\mathcal{H}_1 \otimes \mathcal{H}_2$ is said to be entangled if it is impossible to represent the state as a single product $\ket{a}\otimes\ket{b}$, where $\ket{a} \in \mathcal{H}_1$ and $\ket{b} \in \mathcal{H}_2$. Generalization of tensor product to $n$ number of qubits leads to quantum entanglement of $n$-qubits. In a mutually entangled state of $n >2$ qubits, measurement in any one subsystem reduces partial entanglement that is, the remaining qubits are entangled. The outcomes of local measurement in the next subsystem stay random with no correlation with the first subsystem\,\cite{JJJ2024}. A special class of entangled states that have two properties, (a) the state is in a superposition of all multi-qubit states with the same parity and (b) the probability of each multi-qubit state is equal, is used as a resource for publicly verifiable QRNG (PVQRNG).
The most general four-qubit state satisfying the above two properties is given by  
\begin{equation}
   \begin{split}
      \ket{\Phi_4 ^ {\alpha_1 \cdots \alpha_7}} &= \dfrac{1}{2\sqrt{2}} \bigg ( \ket{0000} +e^{i\alpha_1}\ket{0011}+ e^{i\alpha_2}\ket{0101}\\
      &+ e^{i\alpha_3}\ket{0110}  + e^{i\alpha_4}\ket{1001}+ e^{i\alpha_5}\ket{1010}\\
      &+ e^{i\alpha_1}\ket{1100}+e^{i\alpha_7}\ket{1111} \bigg)  
    \end{split}
\end{equation}
where $\alpha_i \in R $ for parity 0.

\begin{figure}[H]
\centering
\includegraphics[width =\linewidth]{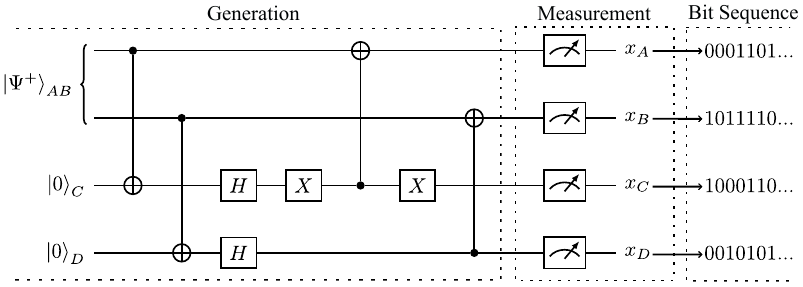}
\caption{Quantum circuit for generating four qubit entangled state using two-qubit entangled state $\ket{\Psi^+}$.  A general output quantum state for this setup is a four-qubit entangled state of the form given in Eq.\,\eqref{eq:theory1}.}
\label{setupcircuit}
\end{figure}

One of the configurations of the quantum circuit to generate a four-qubit entangled state for generating the four-bit string from the measurement outcome is shown in Fig.\,\ref{setupcircuit}. That is, a simple form of a four-qubit entangled state for PVQRNG can be prepared using the sequence of unitary operations on the initial state $|\Psi\rangle_0 = |\Psi^{+}\rangle|00\rangle$ with $\ket{\Psi^+} = \frac{1}{\sqrt{2}} (\ket{\text{01}}_{12} + \ket{\text{10}}_{12})$.

The state generated before the measurement is performed will be,
\begin{equation}\label{eq:theory1}
\begin{split}
\ket{\Psi}_{\text{ABCD}} & =  \frac{1}{2\sqrt{2}} \bigg(\ket{0000} +\ket{0101}+\ket{0110}-\ket{0011} \\
%\text{CNOT}_{CD} \cdot   H_C \cdot \text{CNOT}_{BD} \cdot H_B  \cdot\text{CNOT}_{AD} \cdot H_A \ket{00} \ket{00}  \\
&  +\ket{1100}-\ket{1001}-\ket{1010}-\ket{1111}\bigg ).
\end{split}    
\end{equation}
The above state can be rewritten as 
\begin{equation}
   \begin{split}
          \ket{\Psi}_{\text{ABCD}} &= \dfrac{1}{2\sqrt{2}} \bigg ( \ket{0}_{\text{A}} \otimes \big (\ket{000} +\ket{101}+ \ket{110} -\ket{011} \big )_{\text{BCD}}   \\ 
          & + \ket{1}_{\text{A}}  \otimes \big ( \ket{100}-\ket{001} -\ket{010}-\ket{111} \big )_{\text{BCD}}\bigg).
    \end{split}
\end{equation}
This indicates that in a measurement of any single qubit from the entangled set of four, the remaining three qubits acquire an entangled state of a similar form as the four-qubit. The obtained sequences of output from the measurement on each qubit, though unconditionally random, are correlated via XOR relation,
\begin{equation}
    X_A (i) \oplus X_B (i) \oplus X_C (i) \oplus X_D (i) =0
\end{equation}
for all the permutations of the indices. The quantum entanglement distinguishes them for any set of 4 sequences, where 3 are produced by an independent random number generator and the fourth one by XOR operation. In this case, the randomness of the 3 generated sequences must be proved. This distinguishes our case, where the randomness in each string is intact. The precise statistical coupling in the sequences lies in the symmetry of the $n$-qubit entangled states (for $n > 3$). In simpler terms, when measurements are performed on these entangled states, the resulting binary sequences, while distinct, exhibit identical statistical properties (entropy) \cite{JWW2020}. In the Appendix A,  we have shown the quantum circuit for generation other configuration of four-qubit state for PVQRNG.\\

\noindent
{\it Public randomness verification :} Standard QRNG produces a single sequence of bits which are inherently random, but there exist no method to verify it without accessing the information. However, when we use a four-qubit entangled state of the form given in Eq.\,\eqref{eq:theory1}, the measurement output gives the sequence of bits, $x_A$, $x_B$, $x_C$ and $x_D$. Repeated measurement on identical four-qubit states gives four sequences of output on qubits A, B, C, and D given by $X_A$, $X_B$, $X_C$, and $X_D$, respectively.  These variables are mutually independent pairwise, $I(X_i,X_j) = 0 ,\; \text{for} \; i,j=A,B,C,D $. Therefore, randomness verification of $X_i$ ensures the quality of randomness on the other sequences without explicitly analyzing them. However, the configuration of the four-qubit entanglement ensures that the projective measurement in the computational basis of each subsystem, the random variables for each subsystem will be either 0 or 1, which are denoted as $x_A$, $x_B$, $x_C$, and $x_D$ and obey the relation, 
\begin{equation}
    x_A \oplus x_B \oplus x_C \oplus x_D=0 \label{constraint},
\end{equation}
where $\oplus$ represents the addition modulo 2 operator.
This tells us that the knowledge of any three bits of information helps in the recovery of the fourth bit leaving it vulnerable to privacy of the random bits generated. To address this concern, one of the four sequences of bits generated can be made inaccessible by discarding them, the correlation between the sequences will then be lost and all the other three sequences will be independent of one another. From the three remaining sequences,  public verification of randomness on one sequence ensures the randomness of the other two sequences in real-time without being probed for any direct verification, thus securing the two sequences of random bits. The two sequences can be used as two independent publicly verified secured private random bits or the two can be combined and used as a single sequence of random bits.  We can note that by subjecting $n$ bits for public verification we can obtain $2n$ bit of verified random number. \\

\noindent
{\it Verification of entanglement :}
We should also note that the relation between the measurement outcome given in Eq.\,\eqref{constraint} is a resource for verification of entanglement in the four-qubit state used for QRNG, Eq.\,\eqref{eq:theory1}. If the source is of quantum origin, the condition in Eq.\,\eqref{constraint} along with the successful randomness tests can guarantee the generation of the four-qubit entangled state, in Eq.\,\eqref{eq:theory1}.  For that we need to access all the four bits generated from the measurement outcome. However, instead of accessing all the bits generated, one can sample out some small fraction of measurement outputs validation of Eq.\,\eqref{constraint} can be verified to confirm the entanglement in the four-qubit state used for QRNG. Once the entanglement is verified, the sample output can be discarded and the remaining output can be used for PVQRNG.  Following both procedures will ensure entanglement verification and public verification of the randomness in the bits generated without compromising the security of the bits.  

There is also a possibility of defining a four-qubit entangled state which will enable us to perform real-time entanglement verification and randomness verification by accessing only two of the four random bit sequences.  For that one has to generate a state with a direct correlation between the first two qubits. That state can be generated by removing the last two XOR and X-gates from Fig.\,\ref{setupcircuit} resulting in the circuit as shown in Fig.\,\ref{entcircuit} in Appendix B. The output state will be in the following form, 

\begin{equation}
\begin{split}
    \ket{\Phi}_{ABCD} = \dfrac{1}{2\sqrt{2}} \bigg( \ket{1000}+\ket{0100} -\ket{1010}
+\ket{0110} \\ +\ket{1001} -\ket{0101} -\ket{1011} -\ket{0111}\bigg)
    \end{split}\label{entver}
\end{equation}
and this  can be rewritten as 
\begin{equation}
\begin{split}
    \ket{\Phi}_{ABCD} &= \dfrac{1}{2\sqrt{2}}\bigg [ \big(\ket{01}+\ket{10} \big)_{AB}\otimes \big(\ket{00} - \ket{11}\big)_{CD} \\ &- \big(\ket{10}-\ket{01}\big)_{AB}\otimes \big (\ket{10} - \ket{01}\big )_{CD} \bigg ].
%\\&= \dfrac{1}{2\sqrt{2}} [ (\ket{\text{HV}}+\ket{\text{VH}})\otimes(\ket{00} - \ket{11}) \\ &- (\ket{\text{VH}}-\ket{\text{HV}})\otimes(\ket{10} - \ket{01}) \bigg ].  
\end{split}
\end{equation}
The outcomes from the measurement of the first two qubits will always satisfy
\begin{equation}
    x_A \oplus x_B = 1  \label{constraint2}
\end{equation} 
and the condition given by Eq.\,\eqref{constraint} will not hold.  Thus, the real-time entanglement and randomness verification involves generating four sequences of bits from $n$ copies of the four-qubit entangled state $\ket{\Phi}_{ABCD}$. Each copy is measured in the computational basis, producing strings $X_A$, $X_B$, $X_C$, and $X_D$. The user sends $X_A$ and $X_B$ to a verifier, who performs an XOR continuity test, $X_A(i) \oplus $ $X_B(i) = 1$ and keeps the other two, $X_C(i)$ and $X_D(i)$ hidden and secured. Once the XOR continuity test is verified, one set of the sequence is further sent for randomness verification. Randomness in one ensures randomness in the last two strings because of the mutual independence of the last two random variables.  \\

%=====================

\subsection{Public and Private Key Generation}
\label{PPkey}
%=====================
%\vspace{0.5cm}
%\textit{Public and Private Key Generation}
%\vspace{0.5cm}

Public and private key pairs are essential cryptographic tools for data encryption and decryption. While the public key can be freely distributed without concern, the private key must always be kept secure and never shared unprotected. Here, let us consider two spatially separated parties, Alice and Bob, who share the four-qubit entangled state, Eq.\,\eqref{setupcircuit} between them. We present a method for Alice (Bob) to share an encrypted message with Bob (Alice), which can only be read by Bob (Alice) once Alice (Bob) shares her (his) public key.
Let us say, Alice and Bob simultaneously generate the bits $x_A$, $x_C$(for Alice) and, $x_B$, $x_D$(for Bob).  Alice can access only $x_A$ and $x_C$, while Bob can access only $x_B$ and $x_D$. Alice needs to send a message $m\in\{0,1\}$ to Bob. To encode this message, she performs an XOR operation between the message and her bit $x_A$ resulting in $m\oplus x_A$. She then sends $m\oplus x_A$ along with $x_C$ to Bob. Any third party intercepting $m\oplus x_A$ and $x_C$, cannot deduce the original message $m$ since they lack the necessary information about $x_A$. Bob, on the other hand, can determine $x_A$ upon receiving $x_C$ since Bob knows $x_{B}$, and $x_D$. He can solve for $x_A$ from the relation, $x_A\oplus x_B \oplus x_C \oplus x_D=0$. With $x_A$ known,  Bob can then decode the message $m$ by computing $m=(m\oplus x_A)\oplus x_A$.

In this protocol, the receiver must keep both of his bits private, while the sender must keep one of her bits private and make the other one public to enable the receiver to read the message. Here, $m$ is the message to be communicated, $m\oplus x_{A}$ is the encrypted message protected by Alice's private key, $x_{A}$. Only Bob can decrypt the message once Alice shares her public key, $x_{C}$ using his private keys $x_B$ and $x_D$. Generating a key from entanglement is a well-established concept\cite{KEO2003,IA2004}. However, in past scenarios, the key had to undergo various tests to confirm its randomness after being obtained. In our protocol, we can perform these randomness verification tests on one-bit strings while using other bit strings as keys. This method enables real-time verification and generation, ensuring that the keys we generate are never exposed to randomness testing.

%-================================
\subsection{QBER analysis}
%-================================
\noindent The generated state, $\rho^{'}_{\text{ABCD}}$ may not match the expected state, $\rho_{\text{ABCD}}=\ket{\Psi}_{\text{ABCD}}\bra{\Psi}_{\text{ABCD}}$ from our experimental setup. It is possible that:

\begin{equation}
    \text{F}(\rho^{'}_{\text{ABCD}}, \rho_{\text{ABCD}}) < 1 - \epsilon
\end{equation}
where, F is the fidelity and $\epsilon > 0$.\\\\
The state $\ket{\Psi}_{\text{ABCD}}$ in Eq.\,\eqref{eq:theory1} can also be written as,

 \begin{equation} 
 \ket{\Psi}_{\text{ABCD}} = \dfrac{1}{2\sqrt{2}} \bigg( \ket{0} \otimes  \ket{\Theta_3 ^ {\pi00}} +
    \ket{1} \otimes \ket{\Theta_3^ {'\pi\pi\pi}}\bigg)
\end{equation}
where, $\ket{\Theta_3 ^ {\pi00}} = \dfrac{1}{2}(\ket{000} + e^{i0}\ket{110} +  e^{i0}\ket{101} + e^{i\pi}\ket{011})$ and  $\ket{\Theta_3^ {'\pi\pi\pi}} =\dfrac{1}{2}(\ket{100}+e^{i\pi}\ket{001}+e^{i\pi} \ket{010}+e^{i\pi}\ket{111}) $
Decrease in fidelity results in a deviation from the ideal state. Suppose we perform projective measurement on subsystem A, keeping everything untouched. The projective operators can be defined as $K_A^x = \ket{x}\bra{x}\otimes I_{\text{BCD}}$ for $x \in \{0,1\}$. The state post measurement for system BCD is given by, 
\begin{equation}
    \rho_{BCD}^x = \dfrac{tr_A (K_A^x \rho_{\text{ABCD}} K_A^{x\dagger})}{tr(K_A^{x\dagger}K_A^x \rho_{\text{ABCD}})}.
\end{equation}
The privacy of the state $\rho_{BCD}^0$ can be analyzed as follows.\\
The state $\rho_{BCD}^0$ can be written as,
\begin{equation}
    \rho_{BCD}^0 = (1-p)\ket{\Theta_3 ^ {\pi00}} \bra{\Theta_3 ^ {\pi00}} + \dfrac{p}{8}I_{BCD}
\end{equation}
where $p$ is the probability of getting a maximally mixed state in place of a maximally entangled state. On measuring the system B, C, and D in the computational basis, then there is a probability $p/2$ with which the outcome doesn't match the outcome from $\ket{\Theta_3 ^ {\pi00}}$. Thus, quantum bit error rate (QBER) = $p/2$ for this case.
The privacy of other states like $\rho_{CD}^0$ can also be calculated in a similar way.  \\
The state $\rho_{CD}^{00}$ can be written as,
\begin{equation}
    \rho_{CD}^{00} = (1-p)\ket{\Theta_2 ^ {0\pi}} \bra{\Theta_2 ^ {0\pi}} + \dfrac{p}{4}I_{CD}
\end{equation}
where $\ket{\Theta_2 ^ {0\pi}}$ is the Bell state $\ket{\Phi^-}= \frac{1}{\sqrt{2}}(\ket{00}-\ket{11})$.
On measuring the systems C and D on the computational basis QBER comes out as p/2 for this case as well. It can be readily checked that the QBER remains the same for all other cases ($\ket{\Phi}_{\text{ABCD}}$) as well.

%================================================================================
\section{Experimental realization on a four qubit photonic system}
\begin{figure}[!]
\centering
\includegraphics[width=0.5\textwidth]{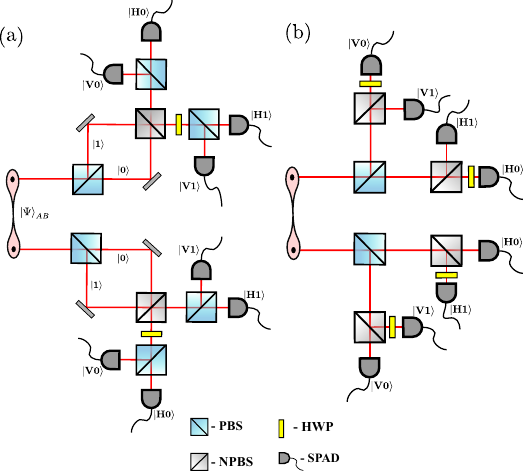}
\caption{The optical scheme for four-qubit state preparation equivalent to quantum circuit given in Fig.\,\ref{setupcircuit} using photon pairs entangled in polarization and path degree of freedom. (a) The scheme involves single-photon interference. (b) The schematic achieves the same quantum state without interference.}
\label{ACT_1}
\end{figure}

\begin{figure*}[t]
\centering
\includegraphics[width = \linewidth]{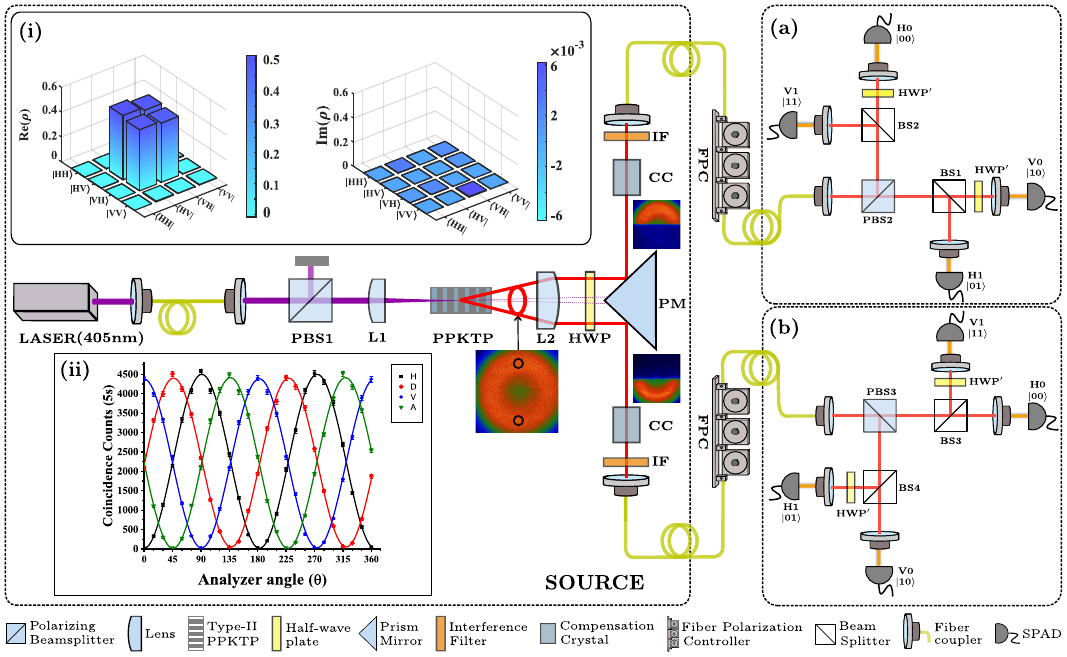}
\caption{Experimental setup: The source part represents the generation of polarized entangled photons through SPDC in a Type-II PPKTP crystal; the generated entangled photons are separated via a prism mirror (PM)and coupled with a fiber. (i) shows the real and imaginary parts of the reconstructed density matrix of the entangled photons with fidelity $98.12\%$, and (ii) Visibility plot in H/V ($98.6\%$) and A/D ($98.1\%$) basis. Fiber polarization controllers (FPC) are used to guarantee that the polarization will remain after going through the fiber. (a) The measurement setup at Alice's side for generating random number sequence $X_A$ and $X_C$. (b) The measurement setup at Bob's side for generating random number sequence $X_B$ and $X_D$.}
\label{setup}
\end{figure*}

%================================================================================
Multiple-qubit quantum state preparation and computation based on these internal degrees of freedom of photons have already been demonstrated \cite{XPAN2018, RWJ2022, CMC2024}. 
A comparable schematic representation of the experimental setup generating a four-qubit state given in Eq.\,\eqref{eq:theory1} using photons entangled in polarization and path degree of freedom is illustrated in Fig.\,\ref{ACT_1}. The experimental realization of the circuit diagram given in Fig.\,\ref{setupcircuit} is shown in Fig.\,\ref{setup}.

To generate a four-qubit state, entangled photon pairs in polarization degree of freedom are used as source. Entangled photons are generated through the type-II SPDC process using a periodically poled KTP crystal. The schematic of the source is illustrated in Fig.\,\ref{setup}. A 10mm long PPKTP crystal was pumped by a continuous-wave diode laser (Surelock, Coherent), having a center wavelength of 405nm. The astigmatism of the laser output pump beam is corrected by coupling with a single-mode optical fiber (SMF). The collimated circular pump beam is allowed to pass through a polarizing beam splitter (PBS1) to set the pump polarization to horizontal. A Plano-convex lens (L1, f = 17.5 cm) is used for focusing the beam at the center of the PPKTP crystal. The crystal temperature is maintained at $32.3\degree C$ for the two down converted photons to have a degenerate wavelength of 810nm. We have selected the SPDC cone angle of approximately 0.83° relative to the pump beam to ensure a comparable counting rate in each spatial mode and to gather roughly twice the number of photon pairs compared to the collinear configuration. The down-converted photons from PPKTP are collimated using a plano-convex lens (L2, f = 10 cm). A prism mirror has been placed for the ease of collecting the signal and idler photons in two separate arms. A half-wave plate (HWP) at $45\degree$ (between the lens and prism mirror), followed by 5mm long KTP crystals is placed as timing compensators in both arms. Bandpass interference filters (IF) of the centre wavelength of $810 \; \text{nm}(\pm \; \text{10nm})$ are used to filter out the residual pump beam and to only allow the down-converted photons. To maintain the polarization, the photons are coupled through the Fiber Polarization Controllers (FPCs)\cite{MCF2005, SHMA2016, SK2023, ACFT2021}. We have generated a bright source of polarization-entangled photons at a rate of $\sim$16\,400 pairs/s/mW. The density matrix of the generated entangled state is determined through quantum state tomography (QST), and the resulting real and imaginary parts are illustrated in the inset of Fig.\ref{setup}. The generated state exhibits a fidelity of $98.12\%$ with the Bell state given by,
\begin{equation}
 \ket{\Psi}_{AB}= \frac{1}{\sqrt{2}}(\ket{H}_{A}\ket{V}_{B}+\ket{V}_{A}\ket{H}_{B})\label{eq:1}.
 \end{equation}
The fitted visibility curves along with the raw data points are also shown in the inset of Fig.\,\ref{setup}. Data were taken at a lower pump power of 0.15mW so that the accidental counts were minimal(16 counts/second). We have obtained visibility of $98.6\%$, and $98.1\%$ in the H/V and A/D basis respectively, and achieved a maximum B-CHSH value(S) violation of $2.77(\pm0.001)$.

Both the entangled photons are directed to the QRNG units (a) and (b) as illustrated in the schematic of the experimental setup. Both the photons entering the QRNG unit have an equal likelihood of being in either the horizontal ($\ket{H}$) or vertical ($\ket{V}$) state. Both the (a) and (b) units consist of a polarizing beam splitter (PBS2, PBS3),
followed by 50:50 non-polarizing beam splitters (BS) in both output paths as illustrated in Fig.\,\ref{setup}. The PBS separates the polarization information while preserving the entanglement and creating two additional path degrees of freedom, and the resultant state would be,
\begin{equation}
     \begin{split}
     \ket{\Psi}_{ABCD}^{I}&= (\textrm{PBS}_{AC}\otimes \textrm{PBS}_{BD})(\ket{\Psi}_{AB}\otimes\ket{0}_{C}\otimes\ket{0}_{D})\\\\
     &\equiv(\textrm{CNOT}_{AC}\otimes \textrm{CNOT}_{BD})(\ket{\Psi}_{AB}\otimes\ket{0}_{C}\otimes\ket{0}_{D})\\\\
     &=\frac{\ket{H0}_{AC}\ket{V1}_{BD}+\ket{V1}_{AC}\ket{H0}_{BD}}{\sqrt{2}}.
     \end{split}
\end{equation}

Here suffixes A, and B represent the polarization information, and C and D represent the path information of the first and second photon respectively. 
Both the outputs from the PBS of the signal photon are guided through non-polarizing beam splitters (BS). When a single photon enters a 50:50 beam splitter, it has an equal probability of being transmitted or reflected. The BS functions similarly to a Hadamard gate, acting on path degrees of freedom to generate the following state: \\

\begin{equation}
     \begin{split}
     \ket{\Psi}_{ABCD}^{II}&= (\textrm{BS}_{C}\otimes \textrm{BS}_{D})\ket{\Psi}_{ABCD}^{I}\equiv (\textrm{H}_{C}\otimes \textrm{H}_{D})\ket{\Psi}_{ABCD}^{I}\\
     \\&=\frac{1}{\sqrt{2}}\left(\ket{H}_{A}\frac{\ket{0}_{C}+\ket{1}_{C}}{\sqrt{2}}\right)\left(\ket{V}_{B}\frac{\ket{0}_{D}-\ket{1}_{D}}{\sqrt{2}}\right) \\\\
      &+\frac{1}{\sqrt{2}}\left(\ket{V}_{A}\frac{\ket{0}_{C}-\ket{1}_{C}}{\sqrt{2}}\right)\left(\ket{H}_{B}\frac{\ket{0}_{D}+\ket{1}_{D}}{\sqrt{2}}\right).\\
   \end{split}
\end{equation}

Now let us introduce half-wave plates at $45\degree$ in the transmitted path($\ket{0}_C$) of the beam splitters in unit (a) and the reflected path ($\ket{1}_D$) of the beam splitters in unit (b) as shown in Fig.\,\ref{setup} to get the final state:
\begin{multline}
     %\begin{split}
     \ket{\Psi}_{ABCD}\\
     \\=\frac{1}{\sqrt{2}}\left(\frac{\ket{V}_{A}\ket{0}_{C}+\ket{H}_{A}\ket{1}_{C}}{\sqrt{2}}\right)\left(\frac{\ket{V}_{B}\ket{0}_{D}-\ket{H}_{B}\ket{1}_{D}}{\sqrt{2}}\right)\\
      \\+\frac{1}{\sqrt{2}}\left(\frac{\ket{H}_{A}\ket{0}_{C}-\ket{V}_{A}\ket{1}_{C}}{\sqrt{2}}\right)\left(\frac{\ket{H}_{B}\ket{0}_{D}+\ket{V}_{B}\ket{1}_{D}}{\sqrt{2}}\right)\\
\\=\frac{\ket{\Phi^+}_{AB}\ket{\Phi^-}_{CD}+\ket{\Psi^-}_{AB}\ket{\Psi^+}_{CD}}{\sqrt{2}}.\\
\end{multline}
where $\ket{\Phi^+}$, $\ket{\Phi^-}$, $\ket{\Psi^+}$, and $\ket{\Psi^-}$ are the four Bell states.
This state can also be written as:
\begin{equation}\label{eq:exp}
\begin{split}
\ket{\Psi}_{ABCD}&=\ket{0}_{A}\left(\frac{\ket{000}+\ket{101}+\ket{110}-\ket{011}}{2\sqrt{2}}\right)_{BCD} \\ 
&+\ket{1}_{A}\left(\frac{\ket{100}-\ket{001}-\ket{010}-\ket{111}}{2\sqrt{2}}\right)_{BCD}\\
 &\equiv \frac{1}{2\sqrt{2}}(\ket{0000} +\ket{0101}+\ket{0110}-\ket{0011} \\ 
 &+\ket{1100}-\ket{1001}-\ket{1010}-\ket{1111})_{ABCD},
\end{split}
\end{equation}
  which is identical to the state given in Eq.\,\eqref{eq:theory1}.  The outputs from the four BS are coupled to multi-mode optical fibers (MMF), which are connected to eight single-photon counting modules (SPCMs). These are connected to a time tagger to record the arrival times of the photons.
  
Ideally, after the entangled photon passes through a PBS on both units (a) and (b), it becomes entangled in the path as well, generating a four-qubit polarization-path entangled state. Applying a Hadamard gate to the path qubit involves interference between the path states. However, we know that interference occurs for photons only if both paths have the same polarization. After passing through the PBS, the polarizations are separated. Therefore, we do not necessarily need a single BS for the paths. Instead, we can use two separate BSs for the two paths and label them accordingly. But in our scheme, on the Alice and Bob sides, at any given time, only horizontal polarisation (H-pol) and vertical polarisation (V-pol) exist. Consequently, the result is unaffected by single-photon interferometer presence or absence. In the end, full polarisation and path information can only be obtained with a polarising beam splitter (PBS), which creates four different paths. To create these four outputs, we employed two beam splitters (BS) in two different paths. We then performed two polarization-dependent actions on these particular labelled pathways. As a result of this labelling, two of the paths are designated as $\ket{0}$ and the remaining two as $\ket{1}$. Finally, we apply the controlled-NOT(CNOT) operation based on our convention. In Fig. \ref{ACT_1}(a), we show the exact optical setup for that state preparation. In Fig. \ref{ACT_1}(b), we show the setup used for our experimental purposes. Both setups lead to the same state preparation.

The state exists in an equal superposition of eight possible output states and satisfies the conditions required for being an entanglement-certified QRNG. Each output from the four beam splitters corresponds to one of these eight states, and their specific outcomes remain entirely uncertain until measurement. Employing a measurement at this stage allows for the generation of four-bit quantum random numbers. Whenever there is a coincidence between one of the detectors from unit (a) and from (b), we record it as a four-bit number. The first and third bits correspond to detection in unit (a) and the second and fourth correspond to detection in unit (b).

To generate the state given by Eq.\,\eqref{entver}, we can remove the last four half-wave plates in the experimental setup given in Fig.\,\ref{setup}. Thus, in the given experimental setup we can generate both the four-qubit state described in the theoretical description to generate real-time publicly verifiable QRNG, for real-time entanglement verification and to use it as a public and private key generation unit.   

%==============
\subsection{Quantum Random Number Generation and Public Verification}
%===========
If we only have access to the strings $X_A$, $X_B$, $X_C$, and $X_D$, then while Eq.\,\eqref{constraint} is a necessary condition, it is not sufficient by itself to prove the faithfulness of the device. In this setup, we generate a source of polarization-entangled photons, which we verify using quantum state tomography, visibility measurements, and CHSH-value measurements. These photons are then passed through beam splitters as described, resulting in the generation of a four-qubit entangled state, as given by Eq.\,\eqref{eq:theory1}. Given the polarization entanglement of the photons, the condition in Eq.\,\eqref{constraint}, together with the randomness of the sequences, should ensure that the output state conforms to Eq.\,\eqref{eq:theory1}. We generated random numbers across a range of pump powers from 0.1 to 6 mW, achieving a collective bit rate of $\sim 60\; \text{kbps/mW}$. At a pump power of 6 mW, we reached a bit rate of $\sim370$ kbps, which can be further increased by boosting the pump power. All the strings were subjected to randomness testing. The sequences are correlated in such a way that they have the same entropy but remain mutually independent \cite{JJJ2024}. We verified that a successful randomness test on one sequence ensures the randomness of the other sequences, as expected.  This randomness test can be conducted by an external party without compromising the secrecy of the other random strings. All the strings $X_A$, $X_B$, $X_C$, and $X_D$ of the generated random numbers passed all 15 NIST tests without any need for post-processing. Each bit $x_i$ is generated at a rate of  $\sim 15$ kbps/mW. Among the four-bit strings, one (e.g., $X_A$) can be used for public randomness testing, while another (e.g., $X_D$) can be discarded for security purposes. The remaining two strings ($X_B$ and $X_C$) can either be used separately or merged to achieve a higher bit rate of 30 kbps/mW for verified and secure random numbers.

\begin{figure}[H]
\includegraphics[width = \linewidth]{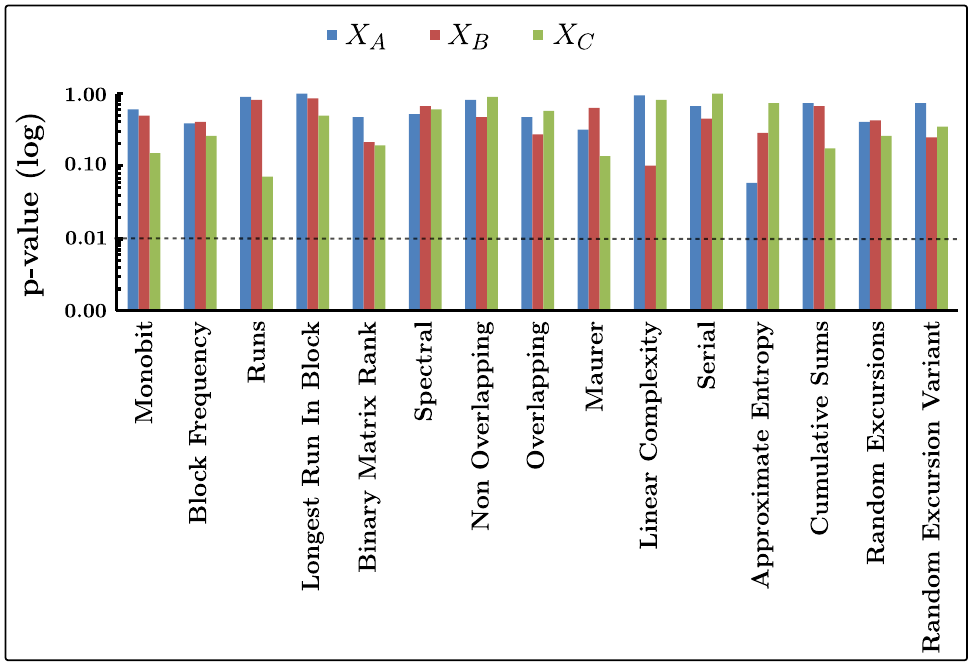}
\caption{NIST randomness test results for the streams of data $X_A$,$X_B$ and $X_C$.  We can note that the data obtained from measurement outcome passes all the NIST tests with a value above 0.01 without the need for any post-processing. }
\label{nist}
\end{figure}

The results of the 15 NIST tests for the data taken at 6mW are shown in Fig.\ref{nist}. The dashed line ($p$ = 0.01) indicates the critical $p$-value above which the sequence is considered random. Results indicate that the statistical patterns are not found in the tested samples of the random bits.

%===============================================

\subsection{Private and Public Key Generation}
Using the setup shown in Fig.\,\ref{setup}, we can achieve a secure communication between two parties as described in Section \ref{Theory}. For this setup, part (a) is assigned to  Alice, and part (b) is assigned to Bob. Alice and Bob can locally and simultaneously generate their respective keys from the coincidence measurement between them and generate keys $x_A$ and $x_C$ for Alice, and $x_B$ and $x_D$ for Bob. To obtain the coincidence measurement and generate keys Alice and Bob have to classically exchange the timing information of photon detection and post-select coinciding events. An eavesdropper cannot find any keys from this timing information since they are not giving out any keys generated corresponding to each detection. The sender can keep one as a private key and announce the other as a public key. Using the information of the public key, the receiver can recover the fourth key described in Sec.\,\ref{PPkey}.

\begin{figure}[H]
\centering
\includegraphics[width=0.5\textwidth]{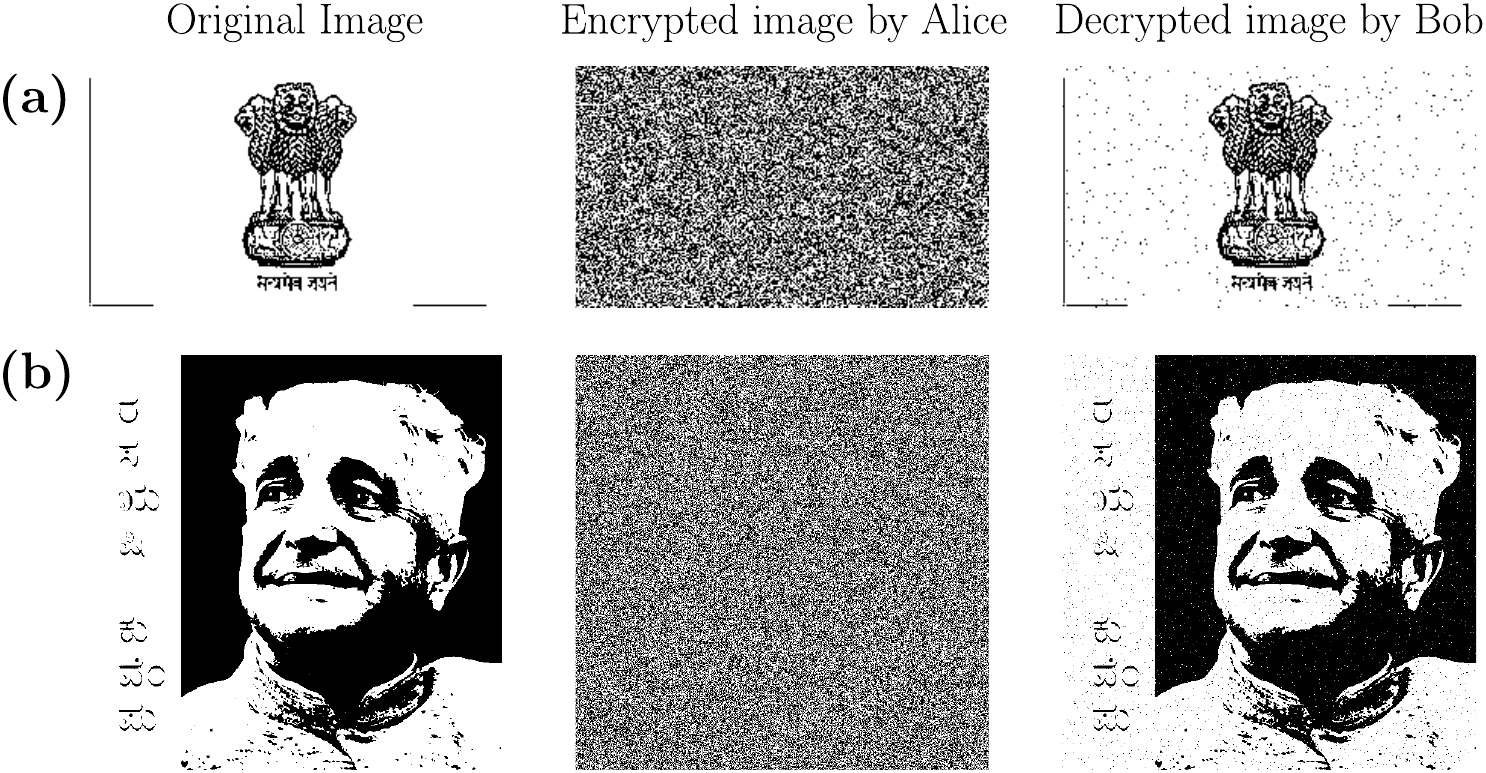}
\caption{Encryption and decryption of image using the private and public key generated from the four-qubit entangled state. (a)Image bit-length is 42\,818 and total key length is 171\,272. (b)Image bit-length is 306\,916 and total key length is 1\,227\,664. }
\label{Q.keys}
\end{figure}

In Fig.\,\ref{Q.keys}, we demonstrate the encryption and decryption of two images using the keys generated from the data we collected. The images used are of bit-length 42\,818 and 306\,916 respectively. For the successful encryption and decryption of these images, the number of four-bit strings ($X_{A}$, $X_{B}$, $X_{C}$, $X_{D}$) should be of the same length as that of the image. So, we have generated keys from the four-qubit entangled state of the same length. This corresponds to a total key length of  171\,272 for image(a) and 1\,227\,664 for image(b) given in Fig.\,\ref{Q.keys}. The string $X_A$(Alice's private key) is used for encrypting the image. For image decryption, the strings $X_B$, $X_C$(Bob's private keys), $X_C$(Alice's public key), and the relation \eqref{constraint} are used. Original, encrypted, and decrypted images are shown in Fig.\,\ref{Q.keys}.

%====================================================
\subsection{Verification of Entanglement}
%====================================================
To verify the polarization entanglement between the two photons from the outputs of the device, the photons from Alice's and Bob's outputs, which exhibit orthogonal polarization, can be separately subjected to visibility or S-value measurements. We ensure that the two photons are polarization-entangled before generating the four-qubit entangled state, and the detailed process is discussed above. Given the polarization entanglement between the two photons, Eq.\,\eqref{constraint} along with the successful randomness tests will guarantee the generation of the four-qubit entangled state given in Eq.\,\eqref{eq:theory1}. 
% \\For verifying the entanglement in the experimental setup generating the four-qubit state of the form given by Eq.\,\eqref{eq:theory1} (with halfwave plates),
We used 10\% of the recorded strings, consisting of 3\,330\,856 bits, to verify if the outputs satisfy the condition in Eq.\,\eqref{constraint}. Of the sample data, 97.9\% met the condition. The sample used for verification can be discarded, allowing the remaining data to be used.

% We have also used the data we obtained from the setup generating state given by Eq.\,\eqref{entver} (after removing the halfwave plates) for real-time entanglement verification. Strings $X_A$ and $X_B$ are used for this purpose and we have observed that 98.2\% of the photons from unit (a) coinciding with the photons from unit B have orthogonal polarization to each other by satisfying the condition Eq.\,\eqref{constraint2}. The rest two strings $X_C$ and $X_D$ are merged together and can be used for cryptographic purposes, with a bit rate of $\sim 30kbps/mW$. 

%========================================
\subsection{Influence of Entanglement Variation and Depolarizing Noise on Four-qubit QRNG system}
%========================================

Variation in the probability amplitudes of the basis state of the four qubit system will result in the variation of entanglement and alter the quality and rate of the verified random number generation. To study the effect of entanglement variation and depolarizing noise on the bitrate of the four-qubit system, we will introduce entanglement control parameters and noise effect only on the first two qubits.  The effect on the first two qubits will reflect on the full four-qubit system. This can be experimentally realized by using the two-photon polarization entangled state,  
\begin{equation}
\ket{\psi}_{\alpha\beta}= \alpha\ket{H}\ket{V}+\beta\ket{V}\ket{H},\label{eq:6}
\end{equation}
 where $|\alpha|^2+|\beta|^2=1$ as the initial state.
 The effective four-qubit state will be, 
\begin{equation}
\begin{split}
\ket{\Psi}_{\text{ABCD}} & =  \frac{\alpha}{2} (\ket{VV00}+\ket{HV10}-\ket{VH01}-\ket{HH11}) \\
&  \frac{\beta}{2}(\ket{HH00}-\ket{VH10}+\ket{HV01}-\ket{VV11}).\\
\end{split}
\end{equation}

\begin{figure}[H]
\centering
\includegraphics[width=0.5\textwidth]{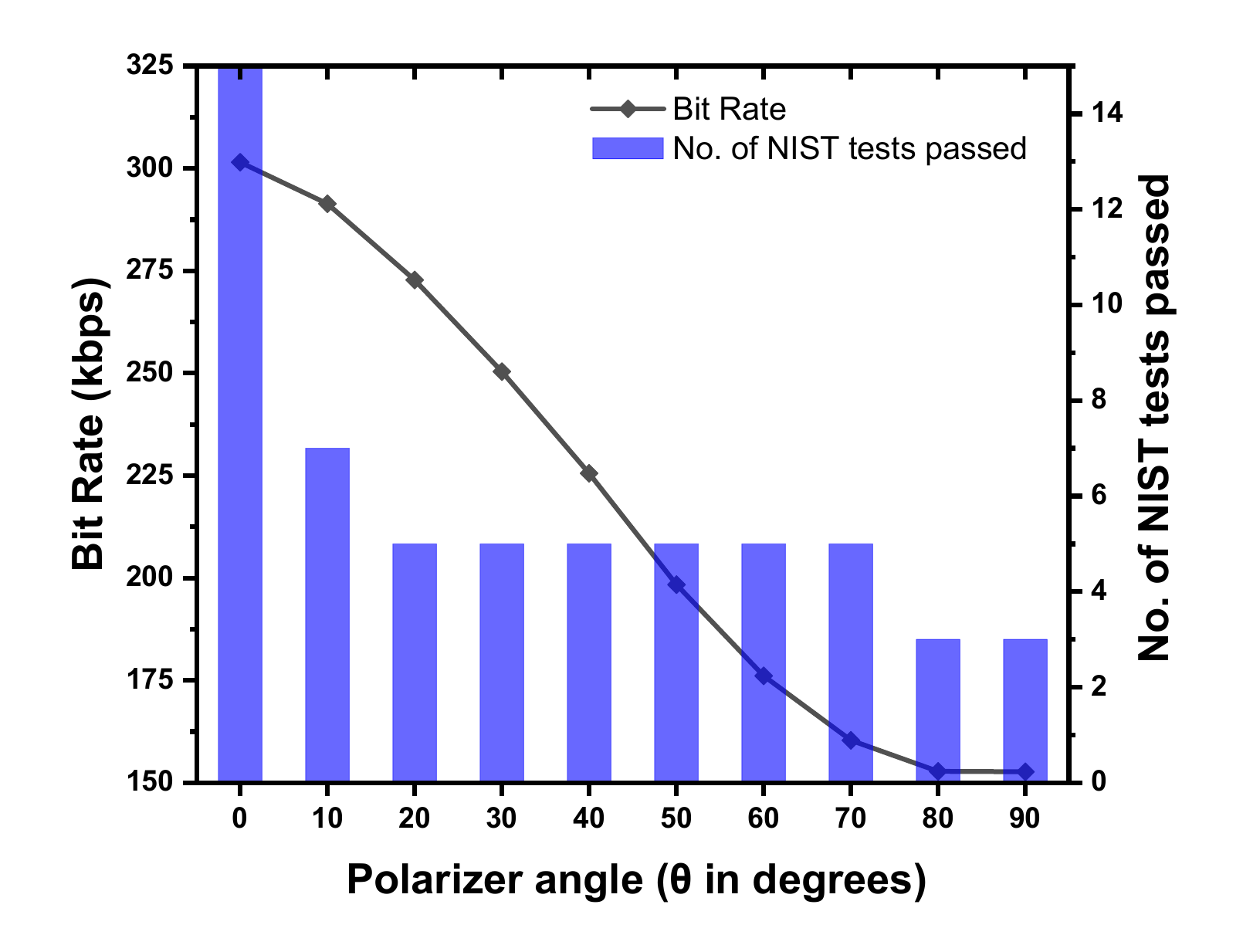}
\caption{Effect of varying degree of entanglement on the bit-rate and the randomness of the bit string used for randomness verification using NIST test (at pump power = 6 mW).  We can note that with a decrease in entanglement, the number of bits generated decreases and the number of NIST tests it passes also significantly goes down.}
\label{noise1}
\end{figure}

When Eq.\,\eqref{eq:6} is the input state, the probabilities of the $\ket{0}$ and $\ket{1}$ outputs from PBS1 would be $\alpha^2$ and $\beta^2$ respectively. That is the probabilities of the generation of the bits 00 and 01 would be $\frac{|\alpha|^2}{2}$ each and that of 10 and 11 would be $\frac{|\beta|^2}{2}$ each, leading to the generation of an unequal number of 0s and 1s in the generated random numbers, depending upon the value of $\alpha$ and $\beta$. As $\alpha$ decreases from $\frac{1}{\sqrt{2}}$ to 0, the state will become less entangled, the difference between the number of 0's and 1's in the generated data increases, and the sequence will become less random.
To experimentally demonstrate this effect, we post-select the photons just before detection by using a polarizer in the transmitted arm of PBS3, varying the number of horizontal photons detected. This manipulation affects the coincidence counts and thus the random numbers generated, mirroring the effects as if the input state is of the form described by Eq.\,\eqref{eq:6}, depending on the polarizer angles. Here, $\alpha^2=\frac{\cos^2\theta}{2}$, where $\theta$ is the polarizer angle. As $\theta$ increases from 0 to 90 (equivalent to $\alpha$ decreasing from $\frac{1}{\sqrt{2}}$ to 0), the generation rate of the sequence and the number of NIST tests the bit string taken for verification passes decrease, as illustrated in Fig.\,\ref{noise1}.  

\begin{figure}[H]
\centering
\includegraphics[width = 0.5\textwidth]{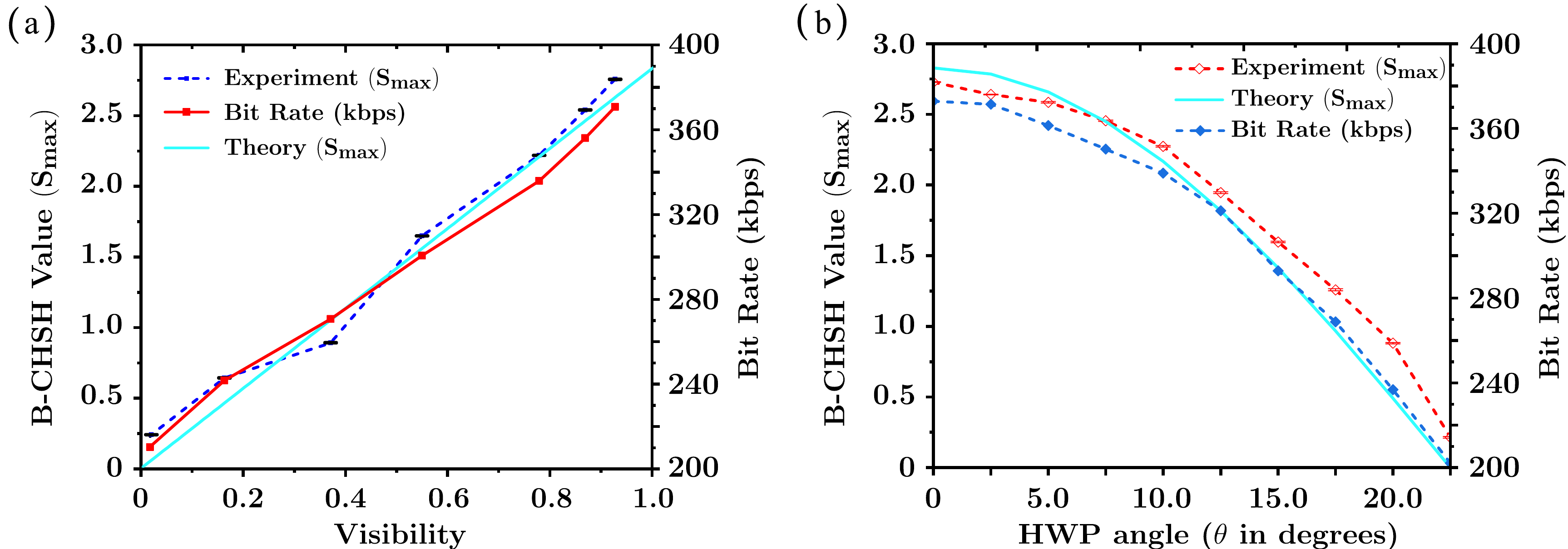}
\caption{Measured bitrate of the generated random numbers and CHSH parameter ($S$ value)  for the corresponding state for different entanglement visibilities (pump power = 6mW). With the decrease in entanglement visibility, we can see a corresponding decrease in $S$ value and secured bit rate generated. The experimental output is in complete agreement with the theoretically expected value.  }
\label{noise2}
\end{figure}

We also reduced the entanglement quality by decreasing the visibilities, achieved through adjustments to the timing compensator. We measured Bell's CHSH value ($S$ value) for the corresponding state and observed its impact on random number generation. The data are depicted in Fig.\,\ref{noise2}(a). 

We can analyze the quantum state using a noisy treatment to theoretically model our experimental data. The noisy state is given by,
\begin{equation}
\rho_{\text{Noisy}} = (1-p)\ket{\Psi_{AB}}\bra{\Psi_{AB}} + \frac{p}{4} \mathbb{I}_{4},
\end{equation}
where \( p = 0 \) corresponds to the pure quantum state, and \( p = 1 \) represents a completely mixed state. To define the visibility, we use the following projection operator,
\begin{equation}
\begin{split}
\ket{\theta_{1}}\otimes\ket{\theta_{2}} = &(\cos{\theta_{1}}\ket{H}_{A} + \sin{\theta_{1}}\ket{V}_{A}) \\
&\otimes (\cos{\theta_{2}}\ket{H}_{B} + \sin{\theta_{2}}\ket{V}_{B}),
\end{split}
\end{equation}
where \(\theta_{1}\) can be fixed to define the basis. For SPDC entangled photons, the A/D (\(\theta_{1} = \pi/4\) or \(3\pi/4\)) basis visibility is crucial for certifying the quality of entanglement. By varying \(\theta_{2}\) and calculating \(\text{Tr}(\rho_{\text{Noisy}} M)\), where 
\begin{equation}
M = \ket{\theta_{1}}\otimes\ket{\theta_{2}}(\ket{\theta_{1}}\otimes\ket{\theta_{2}})^{\dag},
\end{equation}
we obtain a series of values for each \(\theta_{2}\). 

From these values, visibility can be calculated using the formula,
\begin{equation}
V = \frac{v_{\text{max}} - v_{\text{min}}}{v_{\text{max}} + v_{\text{min}}},
\end{equation}
where \(v_{\text{max}}\) and \(v_{\text{min}}\) are the maximum and minimum values obtained, respectively. This procedure is repeated for all \(p\) values. Finally, we will plot Visibility versus $S$-value according to our experimental results.

Implementing depolarizing noise in experiments poses significant challenges. Therefore, we present a mathematical model of depolarizing noise and demonstrate its impact on the generated random numbers.
If we keep a half-wave plate at an angle $\theta$ before PBS3 in the unit (b), state Eq.\,\eqref{eq:1} becomes,
\begin{equation*}
   \begin{split}    
   \ket{\Psi}_{\theta}= \frac{1}{\sqrt{2}}\{(\cos(2\theta)\ket{H}+\sin{2\theta\ket{V}})\ket{V}\\
    +(\sin(2\theta)\ket{H}-\cos{2\theta\ket{V}})\ket{H}\}\\ 
    \end{split}
\end{equation*}
which can also be written as,
\begin{equation}
   \begin{split}
  \ket{\Psi}_{\theta}=\frac{1}{\sqrt{2}}\cos{2\theta}\{\ket{H}\ket{V}-\ket{V}\ket{H}\}\\
    +\frac{1}{\sqrt{2}}\sin{2\theta}\{\ket{H}\ket{H}+\ket{V}\ket{V}\}\\
    =\cos{2\theta}\ket{\Psi^-}+\sin{2\theta}\ket{\Phi^+},
    \end{split}
\end{equation}
where $\ket{\Psi^+}$ and $\ket{\Phi^+}$ are the Bell states. $\ket{\Psi}_{\theta}$ is not maximally entangled in the $H/V$ basis for $\cos{2\theta},\sin{2\theta}\neq0$. As $\theta$ increases from 0 to $\frac{\pi}{8}$, uncertainty in the polarization of the reference photon increases, and at $\frac{\pi}{8}$, it is maximum.

The measured Bell's CHSH value (for a measurement choice, which gives maximal violation for a maximally entangled state on the $H/V$ basis)is depicted against the half-wave plate angles, $\theta\in[0,\frac{\pi}{8}]$ in Fig.\ref{noise2}(b), along with the corresponding reduction in bit rate.

Ideally, measurements of the state $\ket{\Psi}_{ABCD}$ yield uniformly distributed bits $x_A, x_B, x_C,$ and $x_D$, satisfying Eq.\,\eqref{constraint} with maximal Shannon entropy of 1 for each bit. Sampling out a small fraction of measurement output can be used to calculate QBER. For $n$ copies of the entangled state $\ket{\Psi}_{ABCD}$ measured in the computational basis, producing strings $X_A$, $X_B$, $X_C$, and $X_D$ . Assign L = $\{ i : \text{st} \, X_a(i) \oplus X_b(i) \oplus X_c(i) \oplus  X_d(i)  \neq 0\}$, the quantum bit error rate(QBER) is calculated as $\delta = \frac{|L|}{n}$. Similar analysis for  $\ket{\Phi}_{ABCD}$ can be done where $L = \{ i : \text{st} \, X_A(i) \oplus  X_B(i) \neq 1\}$. Imperfections due to non-ideal entanglement or measurements can be mitigated using entanglement purification \cite{PWG2003} and error correction\cite{BHD1996} techniques.

%================================================================================
\section{Conclusion}
In this study, we have demonstrated the practical use of a configurable four-qubit photonic system for quantum-safe applications. Utilizing photon pairs entangled in polarization and path degrees of freedom, our four-qubit system achieves several significant outcomes.  We have reported the generation of publicly verified and secure random bits at a rate of 185 kbps by performing measurements on the four-qubit system and accessing partial information for public verification.  The entangled source photon pairs used for the generation of four-qubit state and randomness generation gives a Bell's-CHSH value of 2.77, and the system maintains a high entanglement verification success rate of $97.9\%$ for the sampled bits from the four-qubit states.  This ensures the secrecy of undisclosed bits and the trustworthiness of the four-qubit system to the beneficiary.   The system simultaneously generates equal numbers of public and private keys, indicating its efficiency in producing secure cryptographic keys essential for quantum communication protocols. This dual-key generation capability enhances the applicability of the system for secure communication infrastructures.
This has been demonstrated by encrypting and decrypting images.

The theoretical model of noise on the four-qubit state and its effect on the generation rate of verified and secure bits aligns with the experimental results. This alignment validates our approach and demonstrates the practical viability of our small-scale multi-qubit photonic system for quantum-safe applications. We also provide a way to verify the security features of the quantum system in real-time, making it easier to integrate into secure communication networks and other applications needing strong quantum security.\\

%================================================================================

\noindent
{\bf  Acknowledgement}\\
We would like to thank Dr. K. Muhammed Shafi for his input on the design of experimental setup and Mayank Joshi for his contribution to build the entangled photon source.  We acknowledge the support from the Office of Principal Scientific Advisor to the Government of India, project no. Prn.SA/QSim/2020.

%============================================================
\appendix
\section{}
\begin{figure}[H]
	\centering
	\includegraphics[width = \linewidth]{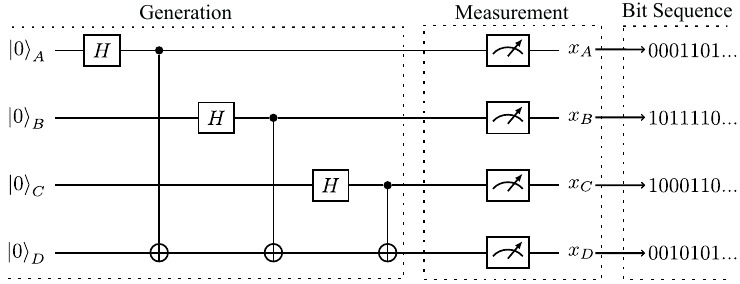}
	\caption{Circuit diagram for generation of alternate four qubit entangles state given in Eq\,\eqref{A} for generation of random bit sequence.}
	\label{gencircuit}
\end{figure}

\noindent 
%=============================================================
The  four-qubit entangled state for PVQRNG can be prepared using the sequence of unitary operations on the initial state $\ket{\Psi}_0 = \ket{0000} = \ket{q}_A\ket{q}_B\ket{q}_C\ket{q}_D$.  In Fig.\,\ref{gencircuit} we have show the operations in in circuit form.\\\\ 

The resulting state after applying the gates as shown in Fig.\,\ref{gencircuit}is:
\begin{equation}\label{A}
	\begin{split}
		\ket{\Psi}_{\text{ABCD}} &= \text{CNOT}_{CD} \cdot H_C \cdot \text{CNOT}_{BD} \cdot \, H_B \cdot \\& 
		\;\;\;\;\;\text{CNOT}_{AD}\cdot H_A \ket{0000} \\
		&= \dfrac{1}{2\sqrt{2}}( \ket{0000} + \ket{0011} + \ket{0101} + \ket{0110} \\
		&\quad + \ket{1001} + \ket{1010} + \ket{1100} + \ket{1111}).
	\end{split}
\end{equation}
The above state can be rewritten as 
\begin{equation}
	\begin{split}
		\ket{\Psi}_{\text{ABCD}} &= \dfrac{1}{2\sqrt{2}} ( \ket{0} \otimes (\ket{000} +\ket{011}+ \ket{101}  + \ket{110}) \\&+ \ket{1} \otimes (\ket{001}+ \ket{010}+ \ket{100}+\ket{111} )).
	\end{split}
\end{equation}

A single copy of the state $\ket{\Psi}_{\text{ABCD}}$ is used to make projective measurements in the computational basis of each subsystem. The random variable for each subsystem is a bit i.e. either 0 or 1 and can be denoted as $x_A$, $x_B$, $x_C$, and $x_D$, as mentioned in the TABLE \,\ref{tab:table1} below.
\begin{table}[H]
	\centering
	\begin{tabular}{c|llll}
		p($x_A,x_B,x_C,x_D$) & $x_A$ & $x_B$ & $x_C$ & $x_D$ \\
		\hline
		1/8 & 0 & 0 & 0 & 0 \\
		1/8 & 0 & 0 & 1 & 1 \\
		1/8 & 0 & 1 & 0 & 1 \\
		1/8 & 0 & 1 & 1 & 0 \\
		1/8 & 1 & 0 & 0 & 1 \\
		1/8 & 1 & 0 & 1 & 0 \\
		1/8 & 1 & 1 & 0 & 0 \\
		1/8 & 1 & 1 & 1 & 1
	\end{tabular}
	\caption{Projection probabilities in the computational basis of subsystems for the output entangled state $\ket{\Psi}_{\text{ABCD}}$ given by Eq \,\eqref{A}. Outcomes with probability zero are not mentioned.}
	\label{tab:table1}
\end{table}

\section{}
\begin{figure}[H]
	\centering
	\includegraphics[width = 0.45\textwidth]{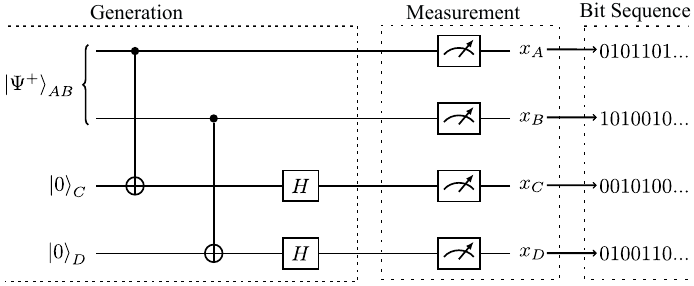}
	\caption{Circuit Diagram for real time entanglment verification of the state $\ket{\Phi}_{\text{ABCD}}$.}
	\label{entcircuit}
\end{figure}

The real time entanglement verification state satisfying Eq\,\eqref{entver} and given by  Eq\,\eqref{constraint2} is generated from the circuit, as shown in Fig.\,\ref{entcircuit}.

\begin{equation}\label{B1}
	\begin{split}
		\ket{\Phi}_{\text{ABCD}} &=  H_C \cdot H_D  \cdot \text{CNOT}_{BD} 
		\cdot \text{CNOT}_{AC} \ket{\Psi^+} \ket{00} \\
		&= \dfrac{1}{2\sqrt{2}} \big( \ket{1000} + \ket{0100} - \ket{1010}
		+ \ket{0110} \\
		&\quad + \ket{1001} - \ket{0101} - \ket{1011} - \ket{0111} \big)
	\end{split}
\end{equation}

with $\ket{\Psi^+} = \frac{1}{\sqrt{2}}(\ket{HV} + \ket{VH})_{\text{AB}}.$ \\\\
Similarly, a single copy of the state $\ket{\Phi}_{\text{ABCD}}$ is used to make projective measurements in the computational basis of each subsystem. The random variable for each subsystem is a bit i.e. either 0 or 1 and can be denoted as $x_A$, $x_B$, $x_C$, and $x_D$, as mentioned in the TABLE \,\ref{tab:table2} below.
\begin{table}[H]
	\centering
	\begin{tabular}{c|lllll}
		p($x_A,x_B,x_C,x_D)$) & $x_A$ & $x_B$ & $x_C$ & $x_D$ \\
		\hline
		1/8 & 1 & 0 & 0 & 0 \\
		1/8 & 0 & 1 & 0 & 0 \\
		1/8 & 1 & 0 & 1 & 0 \\
		1/8 & 0 & 1 & 1 & 0 \\
		1/8 & 1 & 0 & 0 & 1 \\
		1/8 & 0 & 1 & 0 & 1 \\
		1/8 & 1 & 0 & 1 & 1 \\
		1/8 & 0 & 1 & 1 & 1
	\end{tabular}
	\caption{ Projection probabilities in the computational basis of subsystems for the output state $\ket{\Phi}_{\text{ABCD}}$ given by Eq \,\eqref{B1}.}
	\label{tab:table2}
\end{table}
%============================================================================

%========================================================

%===========================================================
\end{document}